# Rapid Design of Top-Performing Metal-Organic Frameworks with Qualitative Representations of Building Blocks

Yigitcan Comlek[1], Thang Duc Pham[2], Randall Q. Snurr[2], Wei Chen[1, *]

[1]Department of Mechanical Engineering, Northwestern University, Evanston, IL, USA

[2]Department of Chemical and Biological Engineering, Northwestern University, Evanston, IL, USA

*Corresponding Author: Dr. Wei Chen

## Abstract

Data-driven materials design often encounters challenges where systems require or possess qualitative (categorical) information. Metal-organic frameworks (MOFs) are an example of such material systems. The representation of MOFs through different building blocks makes it a challenge for designers to incorporate qualitative information into design optimization. Furthermore, the large number of potential building blocks leads to a combinatorial challenge, with millions of possible MOFs that could be explored through time consuming physics-based approaches. In this work, we integrated Latent Variable Gaussian Process (LVGP) and Multi-Objective Batch-Bayesian Optimization (MOBBO) to identify top-performing MOFs adaptively, autonomously, and efficiently without any human intervention. Our approach provides three main advantages: (i) no specific physical descriptors are required and only building blocks that construct the MOFs are used in global optimization through qualitative representations, (ii) the method is application and property independent, and (iii) the latent variable approach provides an interpretable model of qualitative building blocks with physical justification. To demonstrate the effectiveness of our method, we considered a design space with more than 47,000 MOF candidates. By searching only ~1 % of the design space, LVGP-MOBBO was able to identify all MOFs on the Pareto front and more than 97% of the 50 top-performing designs for the $CO_2$ working capacity and $CO_2/N_2$ selectivity properties. Finally, we compared our approach with the Random Forest algorithm and demonstrated its efficiency, interpretability, and robustness.

## Introduction

With recent advances in machine learning (ML), material system design and development has undergone rapid acceleration[1,2]. However, one of the major challenges in applying ML to material system design lies in finding the appropriate design representations[3]. Most material design applications take advantage of quantitative (or numerical) design variables to represent material systems. In most cases, these quantitative descriptors (features) require either expert knowledge or data analysis to find the most appropriate ones. On the other hand, although most qualitative (or categorical) variables (e.g. chemical elements, chemical compositions) are more accessible than quantitative variables, it is challenging to directly include qualitative variables as a part of the design variables in automated materials design. Metal-organic frameworks (MOFs) are an example of such materials systems. MOFs are a class of porous crystalline materials that have been used extensively for gas storage[4,5], gas separation[6,7,8,9], and catalysis[10,11,12]. Because of their highly tunable nature, MOFs have been looked at as a potential solution for different applications such as carbon dioxide ($CO_2$) capture and separation[13,14]. Using a vector notation in which each element corresponds to a qualitative design variable such as topology, node, and edge, MOFs can be represented with the sole usage of qualitative variables as shown in Figure 1. However, the versatility and different possible combinations of the MOF building blocks lead to millions of candidates. To demonstrate a simple example, consider a MOF system with a topology that requires 2 nodes and 3 edges for construction. Selecting only 20 different building blocks for each node and edge leads to a combinatorial design space of more than $10^6$ MOF candidates. Due to the high experimental cost, both in time and resources, computational approaches have been increasingly used to replace experimental exploration[3].

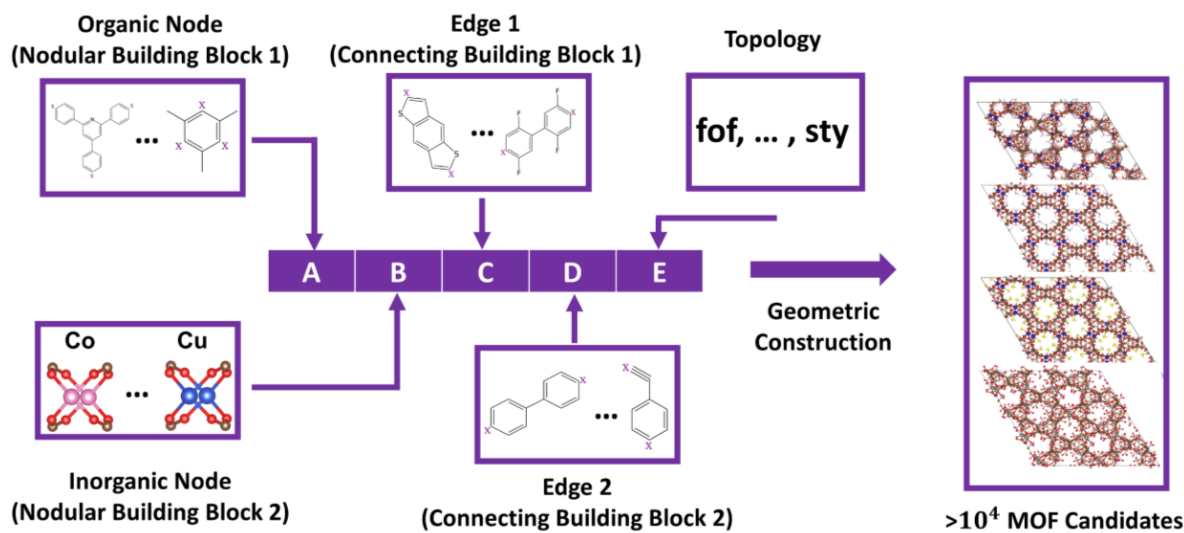

**Figure 1. The qualitative representation and construction of metal-organic framework materials.** Each MOF can be represented with a "vector" where each element (letter) in the vector corresponds to a choice of building block or topology. With the combination of different building blocks and topologies explored in this work, there are more than $10^4$ hypothetical MOFs to be explored.

While high-throughput screening approaches[15,16,17,18,19,20] and ML techniques[21,22,23,24,25,26,27,28] have been utilized to computationally search or design for top-performing MOF structures in different applications, existing approaches usually rely on large data sets and high-dimensional physical descriptors to represent the material design space. These processes can be both time consuming and property specific, meaning that the ML models and descriptors are often not transferable to different design objectives. Finally, many ML models are viewed as 'black boxes' that are not easily interpretable for understanding how and why the model performs the way it is[29,30,31]. Therefore, a novel and a generic computational approach that (i) employs a simple but descriptor-free design representation, (ii) requires substantially smaller amount of data, and (iii) is easily interpretable would be highly useful for the design of novel MOFs.

Bayesian optimization (BO) has been shown to be effective for identifying the optimum candidates for materials systems with large design spaces and local optimums in different applications such as drug discovery, additive manufacturing, and genetics[32,33]. BO has also been used to identify high performance MOFs[34,35]. However, previous works on MOFs require expert knowledge for the choice of appropriate physical descriptors (e.g., gravimetric surface area, largest included sphere diameter) as inputs for surrogate model training. Gaussian Process (GP) is a popular surrogate model choice for BO as it provides both predictions and uncertainty quantification, which are the two main components of the acquisition function for choosing samples when applying BO. However, GP models fall short when there are qualitative design variables. This bottleneck has been recently bypassed by the novel Latent Variable Gaussian Process (LVGP)[36] approach, which can incorporate qualitative variables into GPs by implicitly mapping each qualitative variable into a quantitative space through low-dimensional latent variables. LVGP still possesses the qualities of a GP model in terms of providing fast surrogate modeling, capturing nonlinear responses, providing predictions, and quantifying uncertainties. Furthermore, the latent variable approach provides physics-based dimension reduction. Specifically, the latent variables and their locations in the latent space provide physically meaningful information on how the qualitative variables influence the responses. Thus, LVGP bridges the gap for incorporating qualitative information into engineering design applications and has been already employed in data-driven materials design research[37,38]. Although LVGP and BO have been applied to materials design and development, its application has been limited to qualitative design variables with small number levels, i.e., the design options per variable.

Here we present the Latent Variable Gaussian Process Multi-Objective Batch Bayesian Optimization (LVGP-MOBBO) framework to perform rapid design of superior MOFs directly from the building blocks that construct the material. Specifically, we are interested in identifying the Pareto front for a multi-objective optimization and top-performing MOFs without any human intervention. We are particularly interested in examining the performance of the approach under both small and large numbers of levels for qualitative variables. We take advantage of the readily available qualitative building block information that is used to construct the MOFs and build an interpretable LVGP surrogate model that cooperates with MOBBO to adaptively lead towards promising MOF candidates for $CO_2$ capture and separation. With the integration of batch BO, this work shows that descriptor-free LVGP can also be effectively extended to applications with substantial number of levels.

# Results

**Design Spaces**

To show the effectiveness of LVGP-MOBBO, we demonstrated our framework on a design space using the *fof* topology, which consists of 4 types of building blocks (BB). We used 7 organic nodes (Nodular BB1) and 4 inorganic nodes (Nodular BB2). There are also two types of edges in the *fof* topology, and we used 41 edges for Connecting BB1 and 42 edges for Connecting BB2, as shown in Supplementary Figure 1. For the use of MOFs in carbon capture, two of the most important metrics are the $CO_2$ working capacity and the $CO_2/N_2$ selectivity. Since we focused on method development, we calculated these properties for all MOFs in our design space in advance to aid in testing different variations of the search methods. The first design space, which we denoted as the Reduced Design Space (RDS) for validation purpose, consists of 1001 MOF designs that were specifically selected to demonstrate the applicability of the LVGP and BO onto MOF design problems. The second design space, which we denoted as Entire Design Space (EDS) contains 47,740 MOF design candidates that were constructed by combining all available building blocks (7, 4, 41, 42) for the organic node, inorganic node, and the two edges. Our framework, LVGP-MOBBO, was demonstrated on this design space to show the effectiveness of LVGP when a large number of building blocks (levels) are present.

The two main goals of our design optimization are to identify (i) the Pareto front of MOF designs between the $CO_2$ working capacity and $CO_2/N_2$ selectivity, and (ii) the top-performing MOFs. The Pareto front represents set of MOF designs that possess properties that are superior to the rest of design space but cannot be improved without sacrificing the other properties of interest[39]. Furthermore, the top-performing MOFs represent the MOF designs that are closest to the Utopian MOF design in the Euclidian space. The Utopian MOF corresponds to a hypothetical MOF design that possesses the maximum available property of all objectives, which is often not achievable and therefore considered to be "utopian". As a result, the "top-performing" MOFs are identified based on the distance of a design point to the Utopian MOF in the objective space.

## Framework

We would like to explore a given design space with as few resources as possible. Thus, we implemented the LVGP-MOBBO framework to perform descriptor-free MOF design optimization with only qualitative representations of building blocks. Our proposed LVGP-MOBBO design exploration framework, consists of 5 major parts: Initial Design of Experiments (DOE), Property Evaluation, Latent Variable Gaussian Process, Multi-Objective Batch Bayesian Optimization, and Design Solution (Figure 2).

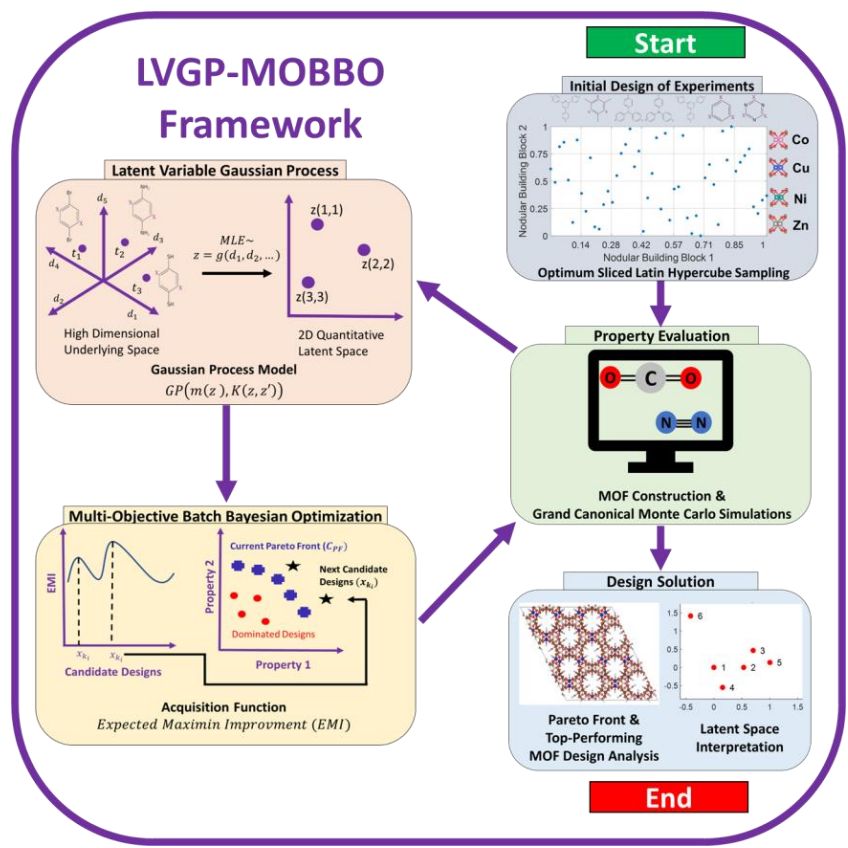

**Figure 2. The Latent Variable Gaussian Process-Multi Objective Batch Bayesian Optimization (LVGP-MOBBO) framework.** The initial set of materials, also known as DOE, is generated by optimal sliced Latin hypercube sampling. Property Evaluation includes MOF construction and prediction of their adsorption properties using Grand canonical Monte Carlo simulations. The LVGP builds the surrogate model that captures the relationship between the design and property space. MOBBO makes the next batch of MOF designs for property optimization. Design Solution analyzes the MOF designs and the latent spaces. The details of each box are explained thoroughly in the *Methods* section.

**Initial Design of Experiments**

For a large design space optimization, the initial selection of design candidates plays a key role. Ideally, they should span the design space as much as possible, for which we employed the optimal sliced Latin hypercube sampling (OSLHS)[40]. The generated MOF designs were then passed into the next task for property evaluation. The detailed generation of the DOE is explained in the *Methods* sections.

**Property Evaluation**

Hypothetical MOFs were created using the ToBaCCo 3.0 package[41], and the geometry optimization was carried out using the LAMMPS code[42] with the UFF4MOF force fields[43]. Grand canonical Monte Carlo (GCMC) simulations were performed using the RASPA package[44] to evaluate the $CO_2$ working capacity and $CO_2/N_2$ selectivity properties. Further details of the property evaluation can be found in the *Methods* section.

**Latent Variable Gaussian Process (LVGP)**

Using the available MOF designs and their associated properties from the GCMC simulations, one LVGP model for each property was trained. Next, the properties of unexplored MOFs in the design space were predicted along with their quantified uncertainty in predictions, which are utilized by the MOBBO. The details of the LVGP modeling are provided in the *Methods* section.

**Multi-Objective Batch Bayesian Optimization (MOBBO)**

Utilizing both the predictions and the uncertainty estimates on the remaining candidates in the design space from the LVGP model, the MOBBO selects a batch of MOFs that has the highest Expected Maximin Improvement (EMI) values. The EMI is formulated in a way that both objectives have equal importance. A batch of *B* number of MOFs designs with the highest EMI values is selected and passed on to the Property Evaluation task once again. The framework then continues in this cycle until the stopping criterion (e.g. number of MOFs identified) is reached. Further details and formulation of the MOBBO are provided under the *Methods* Section.

**Design Solution**

Once the optimization stopping criterion is reached, the identified design candidates are analyzed further to distinguish the Pareto front and the top-performing MOF designs. Finally, the latent space of each building block is visualized to make inferences about their influence on each property of interest.

**Performance**

**Validation using a Reduced Design Space (RDS)**

Before applying our proposed methodology to a large design space, we validated the effectiveness of LVGP and BO on MOFs by implementing the optimization campaign on a relatively small design space. This space contains Connecting BBs that were handpicked to show the novelty of the methodology by validating the latent variables obtained at the end of the optimization campaign and assessing the efficiency of the methodology for designing MOFs that possess superior properties. All the Nodular BBs (7 and 4 levels for Nodular BB1 & BB2) and 6 building blocks from the Connecting BB1 & BB2 were selected for RDS. Specifically, we selected Connecting BBs labeled as {5, 7, 8, 28, 29, 41}. The blocks {5, 7, 8} have similar molecular structures with different functional groups (-CN, -F, -NH$_2$). Blocks {28} and {29} are extended structures of blocks {5} and {7}, respectively. Finally, block {41} is an empty building block, which facilitates a direct connection between the Nodular BB1 and BB2. The building blocks and labels can be found in the Supplementary Figure 1.

The RDS contains 1001 MOF design candidates and three Pareto front MOFs designs. The property space with the known Pareto front and the Utopian designs is shown in Figure 3a. In addition to demonstrating the effectiveness of LVGP for MOFs, the design optimization goal of this study was to identify both the Pareto and other top-performing designs. To account for all the possible Nodular BB1s, 7 MOFs were chosen for the initial DOE using OSLHS. Each of the 7 MOFs corresponds to one level of the Nodular BB1. Furthermore, due to the small number of candidates, we chose to add one MOF design, $B = 1$, during each iteration of BO. Batch BO was implemented on a larger design space, as discussed later in the paper.

The LVGP-BO design optimization campaign ran until the stopping criteria of identifying both the Pareto front and the 10 top-performing MOFs were satisfied. Starting with 7 initial MOFs in the DOE, this stopping criterion led to 44 iterations, which in return shows that a total of 51 (7 + 44) MOFs (5.1% of the design space) are explored as the next design candidates. Specifically, the LVGP-BO framework found all three Pareto front MOF designs in 42 iterations, which corresponds to 4.9% of the design space. The optimization history of identified MOF designs can be seen in Figure 3b. The fast design exploration of the LVGP-BO demonstrated its capability of finding top-performing MOFs.

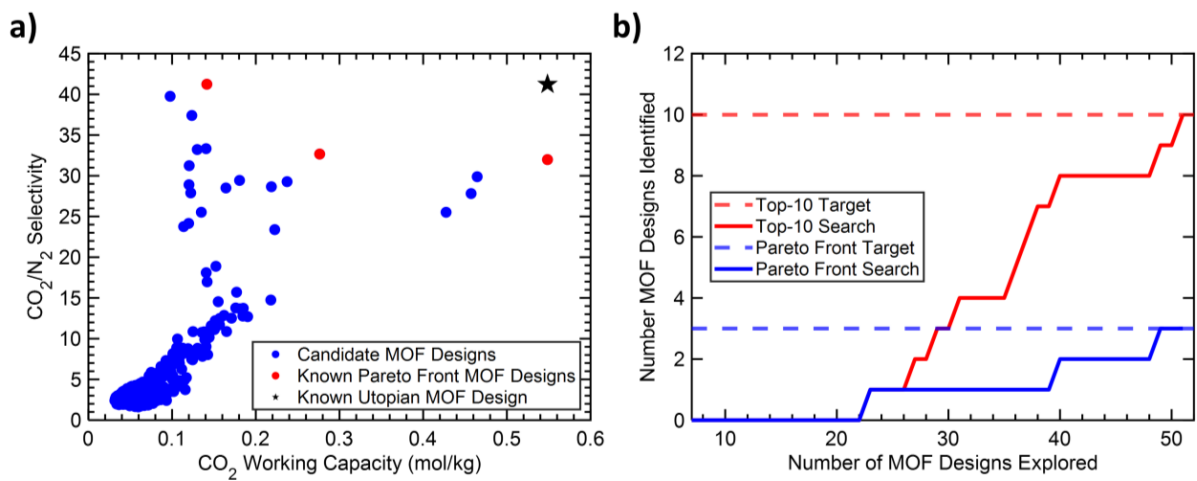

**Figure 3. The LVGP-BO results for the Reduced Design Space (RDS) exploration.** (a) The property space of the available MOF candidates. The known Utopian and Pareto front MOF designs are highlighted with black and red points, respectively. (b) Design optimization history for 10 top-performing and Pareto front MOF designs. The blue color represents the Pareto front search, and the red color represents the 10 top-performing design search.

To verify the interpretability of LVGP models, we examined the latent spaces obtained from training the LVGP surrogate model for both properties after the 44-iteration optimization campaign. In addition, to validate the correctness of the latent variables obtained from the optimization campaign, we also trained the LVGP models on the entire RDS for both properties. We then compared the latent spaces obtained from these two instances. The comparison of the latent spaces for both objectives is shown in Figure 4. The four large colored boxes represent the latent space obtained for each BB after training the LVGP, the two columns represent different properties, and the two rows represent the different training instances. By comparing boxes in each column, we observed that the latent space representations obtained at the end of the design optimization show differences with the latent spaces obtained from the LVGP model trained on the entire design space. This was an expected result since LVGP-BO is optimization driven. However, independent of orientation and the scale of z1 and z2 (the 2D latent space axis), the relative distances between latent variables, which reflect the relationships between design choices (building blocks) and their influence on the properties, are preserved after LVGP-BO. For example, for Connecting BB1, level {41} is far from the other levels for both properties in both training instances in Figure 4. This similarity shows that even though the LVGP used in the design optimization framework was trained on a very small portion of the entire design space that is biased towards promising building block candidates, it can capture the underlying latent variables and the relationships between building blocks very well. This can be very advantageous for designers to understand and extract true meanings from the design decisions that our framework makes.

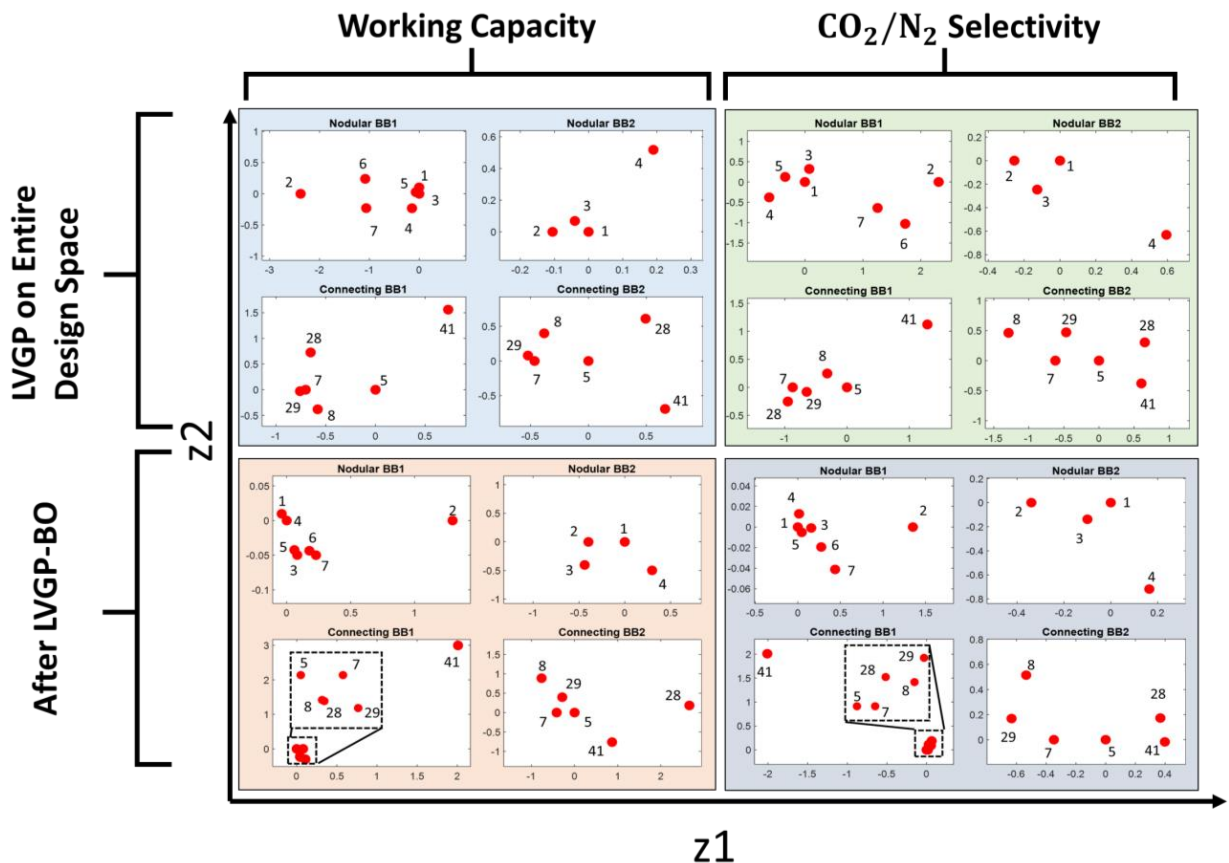

**Figure 4. The latent variables obtained from the RDS study.** Each colored box shows the 4 building block design variables and red dots show their respective latent variable. The numbers represent the design choice for the specific building block. The axes z1 and z2 represent the 2D latent space obtained from the LVGP model. The 1st row represents the latent variables obtained by training LVGP on the entire design space and the 2nd row represents the latent variables obtained after 44 iterations of LVGP-BO. The 1st and 2nd columns represent the $CO_2$ working capacity and $CO_2/N_2$ selectivity properties, respectively. Finally, the dashed boxes show the zoomed in images of clustered latent variables.

The next question then becomes, what do these latent variables represent? Figure 5 shows the importance of the input space, in terms of the textural characteristics, on the property space. For both the RDS and the EDS that will be demonstrated later, we found that most top-performing MOFs for $CO_2/N_2$ separation often have small pores, characterized by low values of the largest cavity diameter (LCD) and small gravimetric surface area (GSA). MOFs with smaller pores could result in stronger van der Waals interaction and thus favor $CO_2$ adsorption over $N_2$ adsorption. Knowing the importance of the input space on the latent space, we further investigated how different building blocks affect the pore size, and ultimately the latent space (Figure 6). For the latent plot of Nodular BB1 (as shown in Figure 4), we found that the distance among the blocks {1, 3, 4, 5} and {6, 7} are small, and block {2} is always far from the rest of the variables. Building blocks {2, 6, 7} are smaller blocks than {1, 3, 4, 5}, resulting in MOFs with smaller LCD (Figure 6a). This could explain why blocks {1, 3, 4, 5} are always closer in the latent space than {2,6,7}. Moreover, block {2} is bulkier, with a branching -$CH_3$ group, than block {6,7}, resulting in MOFs with slightly smaller pores, and thus far away from the other building blocks. In the Connecting BB1 latent variable plots, we observed that the 5 blocks {5, 7, 8, 28, 29} formed a cluster and are located far away from the block {41}. Because {41} is an "empty" building block (Supplementary Figure 1c & d), using block {41} resulted in MOFs with significantly smaller pores than other building blocks (Figure 6b), and thus different in the property space.

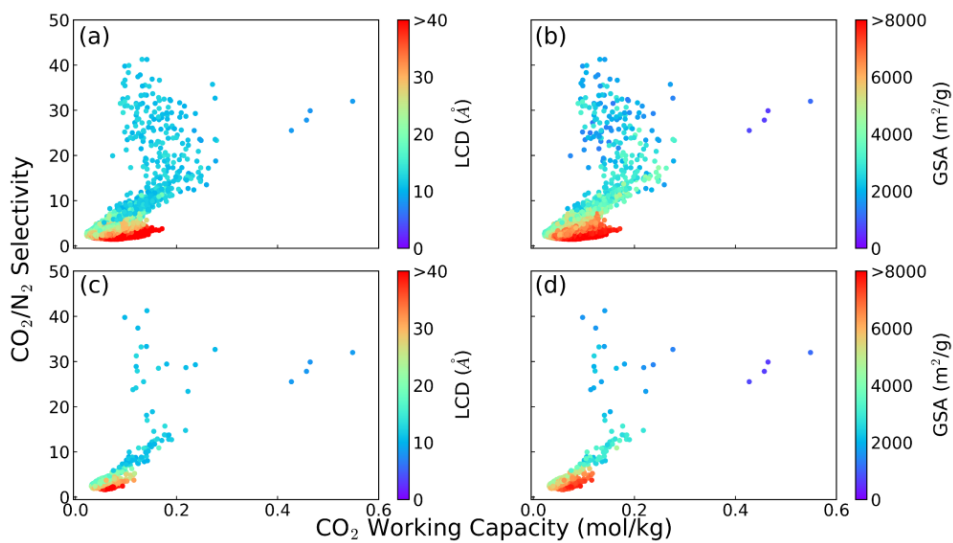

**Figure 5**. **Structure – property relationship of the Entire Design Space (EDS) and Reduced Design Space (RDS) datasets**. The $CO_2/N_2$ selectivity versus the $CO_2$ working capacity for the EDS (a and b) and for the RDS (c and d), colored by the largest cavity diameter (LCD) (a and c) and the gravimetric surface area (GSA) (b and d).

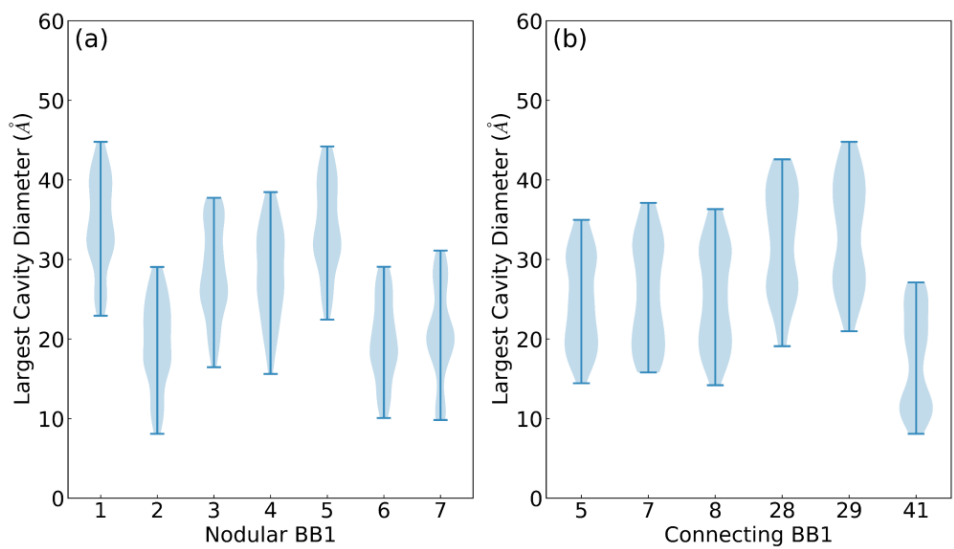

**Figure 6. The distribution of the largest cavity diameters of 1001 MOFs in the RDS for different building blocks.** Largest Cavity Diameter (LCD) distribution for (a) Nodular BB1 and (b) Connecting BB1 on RDS.

For Nodular BB2 and Connecting BB2, we found that the building blocks lead to minimal differences in the pore sizes (Supplementary Figure 2), and thus LCD could not be used to explain the latent space. For Nodular BB2, the building blocks have the same shape and differ only in their metal elements {1: Co, 2: Cu, 3: Ni and 4: Zn}. A potential explanation for the latent space is the difference in Lennard-Jones parameters (Supplementary Table 1), in which Zn has an $\varepsilon$ value of about one order of magnitude larger than the other elements, suggesting a stronger van der Waals interaction for Zn, which could favor $CO_2$ adsorption over $N_2$. As a result, block {4} (or Zn) is far apart from the other designs. Although the chemical identities of the building blocks in Connecting BB2 are the same as in Connecting BB1, Connecting BB2 has small effect on the pore size, and thus the property space. As a result, the points are evenly spread out in the latent space of Connecting BB2. Both the optimization performance and the physically interpretable model obtained from this design study demonstrate the effectiveness of LVGP and BO for further design applications.

**Entire Design Space (EDS)**

After confirming the effectiveness of our methodology on the RDS, we applied our framework to the entire design space (EDS) that contains 47,740 MOF candidates through combination of 7, 4, 41 and 42 building blocks for Nodular BB1, Nodular BB2, Connecting BB1, and Connecting BB2, respectively. The MOF candidates and their respective properties can be seen in Figure 7a. The design space contains 7 Pareto front MOF designs of interest. Incorporating our knowledge from previous LVGP implementations[37,45] and considering the large number of available building blocks in the design space, we decided to match each edge (Connecting BB2, 42 options) with each metal node (Nodular BB2, 4 options), resulting in a total of 168 MOFs to be selected for the initial DOE. To create this DOE, we used OSLHS once again. After starting the LVGP-MOBBO framework with 168 MOFs, we proceeded by adding batches of $B = 5$ new MOFs with highest Expected-Maximin Improvement (EMI) values until the design space search campaign reached the stopping criterion. The design optimization campaign stopped when the mean EMI values of the 5 MOFs that are selected for property evaluation in each iteration is less than a constant, $\delta$, taken as $\delta = 10^{-5}$ in our study.

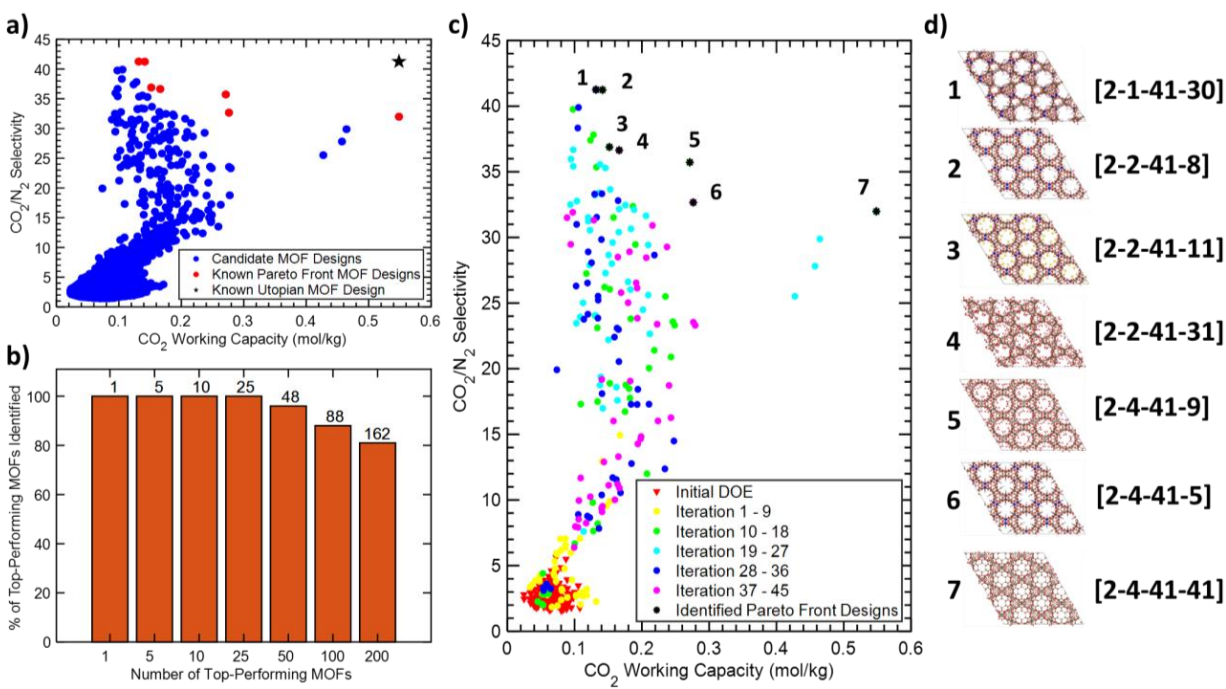

**Figure 7. Performance of the LVGP-MOBBO on the EDS.** (a) The property space of MOF design candidates along with Pareto front and Utopian MOF designs. (b) Percentage of top-performing MOFs identified after 66 iterations. Numbers on top of bars indicate the amount of identified top-performing MOF designs. (c) The initial DOE and the identified MOF designs after different numbers of iterations. (d) The building block representations of Pareto front MOFs and their crystal structures. Each MOF is represented as a vector [A-B-C-D], where each letter represents Nodular BB1, Nodular BB2, Connecting BB1, Connecting BB2, respectively.

With the aforementioned stopping criterion, the LVGP-MOBBO design optimization campaign continued for 66 iterations, identifying 498 MOF designs in total, including the initial 168 MOFs. Our results show that by scanning only 1.04% of the entire design space, LVGP-MOBBO identified all MOF designs that lie on the Pareto front. Specifically, as seen in Figure 7c, all the Pareto front designs are identified within 45 iterations, which corresponds to exploration of only 0.82% of the entire design space. This shows that our methodology is very effective and efficient. Although the initial DOE covers the MOF input design space as evenly as possible, the MOFs in the DOE are not distributed evenly in the property space (Figure 7c). For some machine learning and optimization methods, this can be problematic, as we show later with the Random Forest approach. However, our methodology was swift in guiding the design decisions towards MOFs with high properties. Figure 7b shows the result of exploring the different number of top-performing MOF designs that are closest to the Utopian MOF design. The LVGP-MOBBO found all of the 25 top-performing MOFs. Furthermore, our methodology identified more than 97%, 87%, 80% of the 50, 100, 200 top-performing MOF designs, respectively. Finally, out of all 330 MOFs explored, 206 MOF designs (63.3%) belong to the 330 top-performing MOFs. The high efficiency in identifying a large number of solutions is advantageous due to two potential main reasons. First, it is possible that not every proposed MOF can be synthesized in the laboratory. Second, there are other criteria that must be addressed in practice beyond the $CO_2$ working capacity and selectivity, such as cost and stability. Thus, it is useful to have alternative promising candidates at hand, so that a practical solution can be found.

By looking further into the histogram of selected building blocks at the end of the optimization campaign (Supplementary Figure 3), we observed a bias towards particular building blocks. Specifically, for Nodular BB2 and Connecting BB1, the blocks {2} and {41} are favored because all the Pareto front MOFs possess these building blocks. Therefore, LVGP-MOBBO can identify the promising building blocks effectively and choose them as the next MOF designs at the very early steps of the optimization campaign.

The interpretability of the LVGP approach can be demonstrated using the results for the entire design space. At the end of the 66 iterations, we observed that the latent spaces of the Nodular BB1 (Figure 8a & c) and BB2 (Supplementary Figure 4a & c) converged to a final state. This means that after each iteration, the latent spaces obtained for these design variables did not change. On the other hand, we observed non-convergent latent spaces for Connecting BB1 & BB2 that contain 41 and 42 different design choices, respectively. This is because the LVGP model is trained with a very small percentage of the design space (~1%). The optimization campaign still works well although the latent spaces are not stable. Specifically, block {41} is always separate from the rest in the latent space plots, meaning that its superior effect on the properties is identified clearly. Furthermore, the non-converging behavior is observed for the blocks that have minimal effect on the performance properties. Therefore, our framework can identify the specific building blocks that are superior with a physical justification using the physics-aware LVGP approach.

The latent variable plots of the Nodular BB1 (Figure 8a & c) and Connecting BB1 (Figure 8b & d) can also be explained using the MOF textural properties. For Nodular BB1, blocks {1, 3, 4, 5} form a cluster in the latent space of the $CO_2$ working capacity, while blocks {2, 6, 7} are spread out (Figure 8a). A similar trend was observed in the RDS (Figure 4), which we ascribed to the size of building blocks that determine the LCD of the MOFs. However, the latent space for the $CO_2/N_2$ selectivity changed slightly compared to the RDS. Specifically, blocks {3, 4} are away from blocks {1, 5} and become closer to block {6}, while the positions of the other building blocks remain similar. For Connecting BB1 (Figure 8b & d), block {41} is distant from the other building blocks, which was also observed in the small dataset. Although some building blocks are also further apart from the clusters, their locations change from one iteration to another. The latent variables for the Nodular BB2 and the Connecting BB2 can be found in Supplementary Figure 4. The Nodular BB2 plots can be explained by a similar reasoning as for the RDS plots, whereas the Connecting BB2 is non-convergent due to low training percentage of the LVGP.

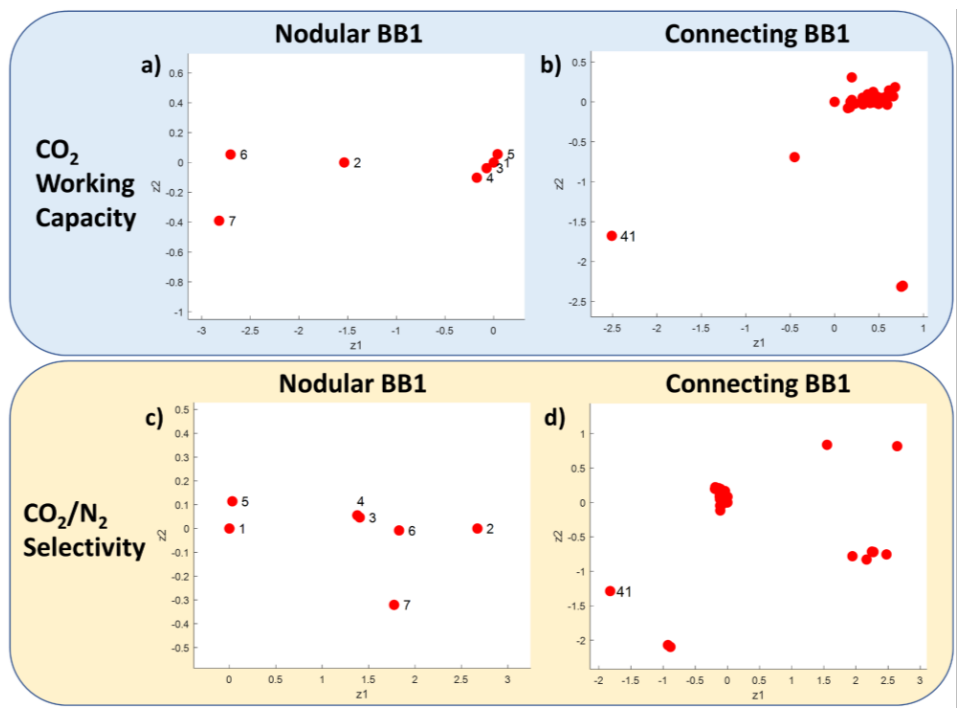

**Figure 8. Latent variable plots after the LVGP-MOBBO campaign on the EDS**. (a) Latent spaces of the Nodular BB1 for (a) the $CO_2$ working capacity and (c) the $CO_2/N_2$ selectivity. Latent spaces of Connecting BB1 for (b) the $CO_2$ working capacity and (d) the $CO_2/N_2$ selectivity.

## Comparison with Random Forest and Robustness of LVGP-MOBBO

We compared our LVGP approach with another ML approach, Random Forest (RF), which is also used for optimization problems with qualitative variables. Both approaches employed the same MOBBO method defined previously. To conduct the study, we ran both frameworks 15 times on the EDS for 60 MOBBO iterations with 15 different initial DOEs. The Pareto front and top-performing MOF design exploration performance of the study can be seen in Figure 9. We observed that the LVGP can identify all the Pareto front MOF designs whereas the RF approach fails to do so in most cases (Figure 9a). The small confidence interval in the performance shows that the LVGP approach is robust and reliable in identifying the Pareto front candidates. On the other hand, the confidence interval of RF is large since some of the RF-MOBBO instances fail to identify any Pareto Front MOF designs. This is because RF-MOBBO is stuck in local optimum designs since the algorithm cannot predict beyond the training data, which usually contained initial DOEs with low properties. In contrast, LVGP was able to expand beyond the low property region towards the high property region by its capabilities of extrapolating beyond training data. Moreover, the Bayesian prediction of uncertainty provided by the LVGP compared to frequentist prediction of RF leads to better and more effective design space exploration[46]. More importantly, the LVGP approach makes the correct design decisions at a faster rate compared to the RF approach, which is crucial if the cost of conducting simulations or experiments is very high. Similarly, for all number of top-performing MOF identification categories, the LVGP approach resulted in a better and more robust performance (Figure 9b).

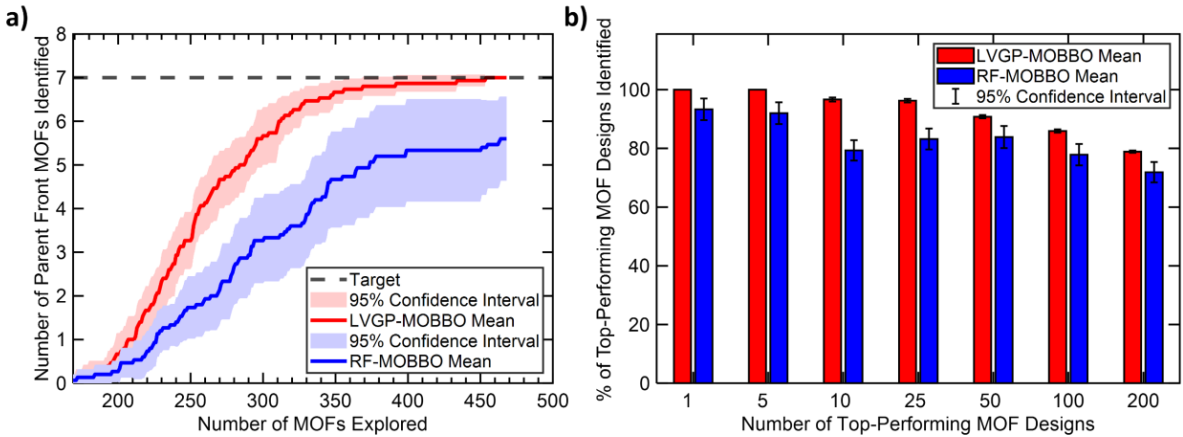

**Figure 9. Comparative Study with Random Forest and LVGP-MOBBO.** (a) Pareto front and (b) top-performing MOF designs identified by the two methods after 15 different runs.

The advantages provided by the LVGP approach are not limited to design optimization. The physical justification provided by the latent variables make the LVGP an interpretable and a favorable ML model for MOF design. The latent variables obtained at the end of the optimization campaign enabled us to gain physical insights behind the design decisions. Although RF is an explainable ML model, model agnostic methodologies are required to draw conclusions or understand its performance. Thus, together with the better performance and accuracy results, the interpretability of LVGP makes our approach more desirable and meaningful for materials design applications.

## Discussion

Due to their versatile and tunable nature, MOFs have very large design spaces, and it is impossible to simulate or perform experiments for every MOF to find the novel candidates for an application of interest. Although numerous ML and high-throughput screening approaches exist, they require either large databases or property-specific descriptors. To tackle these challenges, we presented the LVGP-MOBBO framework to design superior MOFs by only employing qualitative representations of building blocks. The framework presented here provides three main advantages compared to current similar efforts: (i) the framework requires no specific descriptors and only uses the MOF building blocks to perform the adaptive design space search, (ii) the framework is application independent, meaning that it can be applied to any property without the need to select important descriptors for the application of interest, and (iii) the physically justifiable latent variable approach provides interpretability on how each building block influences the resulting performance properties. We demonstrated our framework on a design space with 47,740 MOF candidates. The LVGP-MOBBO successfully identified all Pareto front designs and more than 97% of the 50 top-performing MOF candidate designs by scanning only ~1% of the design space. Compared to Random Forest, LVGP has better performance and robustness, and provides interpretability regarding the design through physically justifiable latent variables. Finally, although we demonstrated our framework on a MOF design space with adsorption properties, LVGP-MOBBO can be applied to any property that requires time consuming simulations such as quantum mechanical calculations.

A key challenge in the presented framework lies in the high number of building blocks. When a large number of blocks are present, although the design optimization campaign works efficiently to identify the top building blocks and MOF designs, the LVGP model struggles to converge to a final latent space due to high number of parameter (latent variables) estimations during model fitting. We expect that by incorporating prior knowledge, when available, into the framework such as assigning prior known distributions to latent variables, the latent variable realizations can be more accurate[47]. We can also incorporate additional descriptors that can further differentiate building blocks from each other to build more accurate LVGP models[48]. An interesting application of our framework would involve performing materials design and development through autonomous experimentation studies. As there is no human intervention in LVGP-MOBBO, and the experimental inputs can be both qualitative and quantitative, we envision that the method we presented here can help researchers guide their experiments efficiently.

## Methods

### MOF Representation and Database Construction

For the *fof* topology, we constructed each MOF using an inorganic node, an organic node, and two edge blocks. Thus, to represent each MOF we use a 5-element 'vector' representation, $[A - B - C - D - E]$, where each letter represents Nodular BB1, Nodular BB2, Connecting BB1, Connecting BB2, and Topology, respectively. A visualization of this representation is shown in Figure 1.

To validate our implementation of LVGP-MOBBO, we created a design space using the *fof* topology, 7 organic nodes (Nodular BB1), 4 inorganic nodes (Nodular BB1), 41 edges (Connecting BB1), and 42 edges (Connecting BB2). From the combinations of the building blocks, we created 48,216 hypothetical MOFs. We eliminated 104 MOFs that had poor initial geometries and missing bonds. We performed geometry optimization on the remaining 48,112 MOFs, in which we found and eliminated 372 MOFs that collapsed after the geometry optimization. Therefore, 47,740 MOFs were considered for this study.

For the initial set of materials, also known as design of experiments (DOE), that initialize the optimization framework, first an optimal Latin hypercube sample with specified number of experiments and variables was created. Then, for each qualitative variable, the design space was sliced into $p_i$ sections, where $p_i$ represents the number of unique options for each qualitative variable. Each DOE design is assigned to a qualitative variable that falls under the sliced section. An example DOE with two qualitative variables (Nodular BB1 & Nodular BB2) that each have 7 and 4 levels is shown in Figure 2 under the *Initial Design of Experiments* box.

## MOF Construction and Geometry Optimization

MOFs were created using the topologically based crystal constructor (ToBaCCo 3.0)[41] software. Geometry optimization was carried out to optimize the unit cell parameters and atomic position using LAMMPS[42] with the UFF4MOF[43] force field. For each structure, the geometry optimization was performed in a cycle that consisted of two steps, as recommended by Anderson et al.[41]. The unit cell parameters and atomic positions were first relaxed using a conjugate gradient (CG) algorithm, followed by atomic position relaxation using the FIRE algorithm (we chose a timestep of 0.1 fs). Each minimization converged only when the change in energy from the previous step to the current step divided by the current energy magnitude was less than $10^{-8}$ and the forces on atoms were less than $10^{-8}$ kcal/mol Å. The cycle stopped when the change in energy between the previous cycle and the current cycle was less than $10^{-8}$ kcal/mol.

## GCMC Simulations

Grand canonical Monte Carlo (GCMC) simulation was carried out using the RASPA package[44]. Each simulation consisted of 10,000 equilibration cycles and 10,000 production cycles. The Monte Carlo moves used were translation, rotation, insertion, deletion, and random reinsertion. Lennard-Jones (LJ) and Coulombic interactions were used to calculate the energies between non-bonded atoms. LJ parameters between different atom types were computed using the Lorentz-Berthelot mixing rules. $CO_2$ and $N_2$ were modeled as three-site rigid molecules with charges on each site, using the LJ parameters and partial charges from the TraPPE force field[49]. LJ parameters for the framework atoms were from the Universal Force Field (UFF)[50]. The framework atom partial charges were calculated using the PACMOF (Partial Atomic Charges in Metal-Organic Frameworks) software[51]. For each MOF, we carried out two GCMC simulations; the first was at the adsorption condition of 1 bar, 313 K, and a bulk molar composition of $CO_2 : N_2 = 0.15 : 0.85$, and the second was at the desorption condition of 0.1 bar, 313 K, and a bulk molar composition of $CO_2 : N_2 = 0.9 : 0.1$.

We used the $CO_2$ working capacity ($\Delta N_{CO_2}$) and the $CO_2/N_2$ selectivity at adsorption ($\alpha_{CO_2/N_2}^{ads}$) as the criteria to determine top-performing MOFs for $CO_2/N_2$ separation. The two properties are defined as follows:

$$\Delta N_{CO_2} = N_{CO_2}^{ads} - N_{CO_2}^{des} \quad (1)$$

$$\alpha_{CO_2/N_2}^{ads} = \frac{N_{CO_2}^{ads}}{N_{N_2}^{ads}} \frac{y_{N_2}^{ads}}{y_{CO_2}^{ads}} \quad (2)$$

Here, $\Delta N_{CO_2}$ is the $CO_2$ working capacity, $N_{CO_2}^{ads}$ and $N_{CO_2}^{des}$ are the $CO_2$ adsorption loadings at the adsorption and desorption conditions, $\alpha_{CO_2/N_2}^{ads}$ is the selectivity of $CO_2$ over $N_2$ at adsorption condition, $N_{N_2}^{ads}$ is the $N_2$ loading at adsorption, and $y_{N_2}^{ads}$ and $y_{CO_2}^{ads}$ are the bulk mole fractions of $N_2$ and $CO_2$ at adsorption, respectively. While $CO_2$ working capacity reflects how effective the MOF is at both capturing and releasing $CO_2$, the selectivity determines how selectively the MOF can separate $CO_2$ from the mixture of $CO_2$ and $N_2$.

**Latent Variable Gaussian Process (LVGP)**

One of the main contributions of this paper lies in the design optimization of MOFs using only the readily available qualitative representations of building blocks. On the other hand, due to the nature of the correlation functions, it is not possible to directly implement the building blocks into the Gaussian Process (GP) models as the difference between variables becomes unclear. Therefore, in this paper, we implemented the novel Latent Variable Gaussian Process (LVGP) to account for the qualitative variables in the GP model[36]. It is known that for every qualitative variable, there are underlying, possibly high-dimensional, quantitative variables that explain its effect on properties. The latent variable approach helps us to map the qualitative variables to a quantitative latent space. Consider a GP model input with $t = [t_1^q, t_2^q, \ldots, t_n^q] \in R^{q \times n}$ with $n$ qualitative variables and $q$ number of points. Each variable, $t_i$, has $p_i$ unique levels (design choices) $\{l_1(t_i), l_2(t_i), \ldots l_{p_i}(t_i)\}$ for $i = 1{:}n$, (e.g., Cu, Co, Ni options for nodular building block ($p_i = 3$)). Then, each qualitative variable can be represented with a latent variable vector $z_t(l_{p_i}) = \{z_t^i(l_1), \ldots, z_t^i(l_{p_i})\}$ for $z^i \in R^k$, where k represents the dimension of the $z^i$. The developers of the algorithm have stated that users are free to choose the dimensions of $z^i$ but also demonstrated that $k = 2$ is enough to represent the underlying high-dimensional quantitative space. Consequently, we chose $k = 2$. Thus, each level within a qualitive variable can be represented with a latent vector of $z_t(l_{p_i}) = \{z_t^1(l_{p_i}), z_t^2(l_{p_i})\}$, and the input to the GP model becomes $z(t) = [z_t^1(l_{p_i}), z_t^2(l_{p_i}), \ldots, z_t^1(l_{p_n}), z_t^2(l_{p_n})] \in R^{q \times 2n}$. An illustration of the latent variable representation of qualitative building blocks is shown in Figure 2 under the *Latent Variable Gaussian Process* box. Consider a typical single response Gaussian Process model, which consists of prior constant mean $\mu$ and $K_Z(t)$, describing the mean response at any given point in the input space, and a zero-mean Gaussian Process with a covariance function $K(t, t')$, respectively. The covariance function $K(t, t')$ determines the relationship or the correlation between variables in the model. The covariance function can be further extended to $K(t, t') = \sigma^2 \cdot c(t, t')$ where the $\sigma^2$ represents the prior variance of the GP model and $c(t, t')$ describes the correlation between each point in the model through the specified correlation function. To explain the relationship between each design candidate for this application, we have implemented the Gaussian correlation function:

$$c(t, t') = \exp\left\{-\sum_{i=1}^{p}(\|z_t^1(l_{p_i}) - z_{t'}^1(l_{p_i})\|_2^2 + \|z_t^2(l_{p_i}) - z_{t'}^2(l_{p_i})\|_2^2)\right\} \quad (3)$$

The Gaussian correlation function shown in equation (3) evaluates the correlation between points $t$ and $t'$ based on 2-norm distance. The main reason behind choosing this correlation function is because we assume that points that are close in the spatial input space should also reflect a similar behavior in the output space as well. Along with the 2D mapped latent variables $z = \left(z_t^1(l), z_t^2(l)\right)$ for level $l$ of each qualitative variable $t$, the parameters, $\mu$ and $\sigma$ are estimated through Maximum Likelihood Estimation (MLE) of the log-likelihood function

$$l(\mu, \sigma, z) = -\frac{q}{2}\ln(\sigma) - \frac{1}{2}\ln|C(z)| - \frac{1}{2\sigma^2}(y - \mu\mathbf{1})^T C(z)^{-1}(y - \mu\mathbf{1}) \quad (4)$$

where $q$ is the number of samples, $C$ is the $q \times q$ correlation matrix with $C_{ij} = c(t^i, t^j)$ for $i,j = 1,2,3,\ldots,q$, $\mathbf{1}$ is a vector of ones with dimensions of $q \times 1$, and $y$ is the observed response with dimensions of $q \times 1$. Finally, the 2D quantitative latent variables are then used to construct a GP model that provides both prediction and statistical representation of uncertainty in the design space for Bayesian optimization.

## Multi-Objective Batch Bayesian Optimization (MOBBO)

Bayesian Optimization is a well-known efficient, fast, and easy-to-implement optimization technique that has been used in numerous materials design applications. For single objective optimization, BO makes the decision on which design in the design space should be sampled next based on the choice of an acquisition function. Three well-known acquisition functions are Lower Confidence Bound[52], Probability of Improvement[53], and Expected Improvement (EI)[54]. With its ability to balance exploration of the design space and exploitation of the objective, EI has been a popular choice for most materials design applications. Considering the large MOF design space, we have also chosen EI as our base acquisition function. For a given candidate design $x$, with its predicted objective value $y'_x$ and quantified uncertainty $\sigma_x$ from the LVGP model, EI for single objective optimization can be calculated using,

$$EI(x) = (y^* - y'_x) \cdot \psi\left(\frac{(y^* - y'_x)}{\sigma_x}\right) + \sigma_{x_k} \cdot \phi\left(\frac{(y^* - y'_x)}{\sigma_x}\right) \quad (5)$$

where $y^*$ is the best observed objective so far in the optimization campaign and $\psi, \phi$ represent the cumulative distribution function (CDF) and probability distribution function (PDF), respectively. As equation (5) shows, the EI function suggests a new design by not only considering the exploitation of the objective function, $(y^* - y'_x)$, but also the uncertainties associated with the design space, $\sigma_x$.

Often there are tradeoffs between objectives, meaning that one objective cannot be optimized without sacrificing the other one. This type of problem is also known as Pareto front optimization and is frequently observed in material systems[39]. Thus, for multi-objective optimization problems, the goal becomes discovering the Pareto front of the property space. Therefore, we have expanded the EI formulation by implementing the Expected-Maximin Improvement (EMI) acquisition function to serve as the balancer of the exploration and exploitation for multi-objective optimization. For the case of optimizing two objectives, the formulation of EMI is

$$EMI(x_k) = \min(\max(EI_1, EI_2), 0) \quad (6)$$

where $EI_j$ corresponds to the Expected Improvement value of each objective $j$. The EI formula was used to compare the candidates with respect to the observed number of $p$ Pareto front designs so far in the optimization campaign. Therefore, each $EI_j$ is a vector of $p \times 1$ that contains the EI values of a candidate design on the observed Pareto front designs for each objective. Lastly, we first take the maximum of EI's for both objectives to observe the dominance of the candidate on the current Pareto frontier and then select the minimum of the maximum EIs to balance the multi-objective search. As a result, the EMI is formulated in a way that both objectives have equal importance. Equation (6) selects the single best candidate in each multi-objective BO iteration.

Due to the large number of candidate designs and the cost of training GP models, it is not ideal to train the LVGP with a single design candidate at each iteration. Therefore, to extend single candidate BO to select a batch of promising candidates, we select $B$ candidates that possess the highest EMI values in each iteration and use them as the next design candidates. A demonstration of a single MOBBO iteration is demonstrated in Figure 2 under the *Multi-Objective Batch Bayesian Optimization* box.

# Data Availability

The crystal structures and calculated properties of 47,740 MOFs in this study are deposited on Zenodo. (https://zenodo.org/record/7539659#.Y8Xcr9LMK0o).

# Code Availability

The LVGP-MOBBO code used to carry out this work are described in the Methods section. The LVGP-code can be accessed through the Comprehensive R Archive Network (CRAN) at https://cran.r-project.org/web/packages/LVGP/index.html and MOBBO code is available from the corresponding author upon request.

# References

1. Moosavi, S. M., Jablonka, K. M. & Smit, B. The role of machine learning in the understanding and design of materials. *J. Am. Chem. Soc.* **142**, 20273-20287 (2020).
2. Ramprasad, R., Batra, R., Pilania, G., Mannodi-Kanakkithodi, A. & Kim, C. Machine learning in materials informatics: Recent applications and prospects. *npj Comput. Mater.* **3**, 54 (2017).
3. Jablonka, K. M., Ongari, D., Moosavi, S. M. & Smit, B. Big-data science in porous materials: Materials genomics and machine learning. *Chem. Rev.* **120**, 8066-8129 (2020).
4. Suh, M. P., Park, H. J., Prasad, T. K. & Lim, D.-W. Hydrogen storage in metal–organic frameworks. *Chem. Rev.* **112**, 782-835 (2012).
5. He, Y., Zhou, W., Qian, G. & Chen, B. Methane storage in metal–organic frameworks. *Chem. Soc. Rev.* **43**, 5657-5678 (2014).
6. Li, H. *et al.* Recent advances in gas storage and separation using metal–organic frameworks. *Mater. Today* **21**, 108-121 (2018).
7. Shah, M., McCarthy, M. C., Sachdeva, S., Lee, A. K. & Jeong, H.-K. Current status of metal–organic framework membranes for gas separations: Promises and challenges. *Ind. Eng. Chem. Res.* **51**, 2179-2199 (2012).
8. Roohollahi, H., Zeinalzadeh, H. & Kazemian, H. Recent advances in adsorption and separation of methane and carbon dioxide greenhouse gases using metal–organic framework-based composites. *Ind. Eng. Chem. Res.* **61**, 10555-10586 (2022).
9. Li, J.-R., Kuppler, R. J. & Zhou, H.-C. Selective gas adsorption and separation in metal–organic frameworks. *Chem. Soc. Rev.* **38**, 1477-1504 (2009).
10. Wang, Q. & Astruc, D. State of the art and prospects in metal–organic framework (mof)-based and mof-derived nanocatalysis. *Chem. Rev.* **120**, 1438-1511 (2020).
11. Wei, Y.-S., Zhang, M., Zou, R. & Xu, Q. Metal–organic framework-based catalysts with single metal sites. *Chem. Rev.* **120**, 12089-12174 (2020).
12. Freund, R. *et al.* The current status of mof and cof applications. *Angew. Chem. Int. Ed.* **60**, 23975-24001 (2021).
13. Li, J.-R. *et al.* Carbon dioxide capture-related gas adsorption and separation in metal-organic frameworks. *Coord. Chem. Rev.* **255**, 1791-1823 (2011).

14	Sumida, K. *et al.* Carbon dioxide capture in metal–organic frameworks. *Chem. Rev.* **112**, 724-781 (2012).
15	Avci, G., Velioglu, S. & Keskin, S. High-throughput screening of mof adsorbents and membranes for h2 purification and co2 capture. *ACS Appl. Mater. Interfaces* **10**, 33693-33706 (2018).
16	Altintas, C. *et al.* An extensive comparative analysis of two mof databases: High-throughput screening of computation-ready mofs for ch4 and h2 adsorption. *J. Mater. Chem. A* **7**, 9593-9608 (2019).
17	Gu, C., Yu, Z., Liu, J. & Sholl, D. S. Construction of an anion-pillared mof database and the screening of mofs suitable for xe/kr separation. *ACS Appl. Mater. Interfaces* **13**, 11039-11049 (2021).
18	Li, S., Chung, Y. G. & Snurr, R. Q. High-throughput screening of metal–organic frameworks for co2 capture in the presence of water. *Langmuir* **32**, 10368-10376 (2016).
19	Colón, Y. J. & Snurr, R. Q. High-throughput computational screening of metal–organic frameworks. *Chem. Soc. Rev.* **43**, 5735-5749 (2014).
20	Islamov, M. *et al.* High-throughput screening of hypothetical metal-organic frameworks for thermal conductivity. *npj Comput. Mater.* **9**, 11 (2023).
21	Lee, S. *et al.* Computational screening of trillions of metal–organic frameworks for high-performance methane storage. *ACS Appl. Mater. Interfaces* **13**, 23647-23654 (2021).
22	Park, J., Lim, Y., Lee, S. & Kim, J. Computational design of metal–organic frameworks with unprecedented high hydrogen working capacity and high synthesizability. *Chemistry of Materials* **35**, 9-16 (2023).
23	Yao, Z. *et al.* Inverse design of nanoporous crystalline reticular materials with deep generative models. *Nat. Mach. Intell.* **3**, 76-86 (2021).
24	Chong, S., Lee, S., Kim, B. & Kim, J. Applications of machine learning in metal-organic frameworks. *Coord. Chem. Rev.* **423**, 213487 (2020).
25	Li, Z. *et al.* Machine learning using host/guest energy histograms to predict adsorption in metal–organic frameworks: Application to short alkanes and xe/kr mixtures. *J.Chem.Phys.* **155**, 014701 (2021).
26	Sun, Y. *et al.* Fingerprinting diverse nanoporous materials for optimal hydrogen storage conditions using meta-learning. *Sci. Adv.* **7**, eabg3983
27	Fernandez, M., Boyd, P. G., Daff, T. D., Aghaji, M. Z. & Woo, T. K. Rapid and accurate machine learning recognition of high performing metal organic frameworks for co2 capture. *J. Phys. Chem. Lett.* **5**, 3056-3060 (2014).
28	Shi, Z. *et al.* Techno-economic analysis of metal–organic frameworks for adsorption heat pumps/chillers: From directional computational screening, machine learning to experiment. *J. Mater. Chem. A* **9**, 7656-7666 (2021).
29	Rudin, C. Stop explaining black box machine learning models for high stakes decisions and use interpretable models instead. *Nat. Mach. Intell.* **1**, 206-215 (2019).
30	Wei, J. *et al.* Machine learning in materials science. *InfoMat* **1**, 338-358 (2019).
31	Guo, K., Yang, Z., Yu, C.-H. & Buehler, M. J. Artificial intelligence and machine learning in design of mechanical materials. *Mater. Horiz.* **8**, 1153-1172 (2021).
32	Wang, K. & Dowling, A. W. Bayesian optimization for chemical products and functional materials. *Curr. Opin. Chem. Eng.* **36**, 100728 (2022).
33	Frazier, P. I. & Wang, J. Bayesian optimization for materials design in *Information science for materials discovery and design* (eds Turab Lookman, Francis J. Alexander, & Krishna Rajan) 45-75 (Springer International Publishing, 2016).
34	Taw, E. & Neaton, J. B. Accelerated discovery of ch4 uptake capacity metal–organic frameworks using bayesian optimization. *Adv. Theor. Simul.* **5**, 2100515 (2022).
35	Deshwal, A., Simon, C. M. & Doppa, J. R. Bayesian optimization of nanoporous materials. *Mol. Syst. Des. Eng.* **6**, 1066-1086 (2021).

36  Zhang, Y., Tao, S., Chen, W. & Apley, D. W. A latent variable approach to gaussian process modeling with qualitative and quantitative factors. *Technometrics* **62**, 291-302 (2020).
37  Zhang, Y., Apley, D. W. & Chen, W. Bayesian optimization for materials design with mixed quantitative and qualitative variables. *Sci. Rep.* **10**, 4924 (2020).
38  Iyer, A. *et al.* Data centric nanocomposites design via mixed-variable bayesian optimization. *Mol. Syst. Des. Eng.* **5**, 1376-1390 (2020).
39  Censor, Y. Pareto optimality in multiobjective problems. *Appl. Math. Optim.* **4**, 41-59 (1977).
40  Ba, S., Myers, W. R. & Brenneman, W. A. Optimal sliced latin hypercube designs. *Technometrics* **57**, 479-487 (2015).
41  Anderson, R. & Gómez-Gualdrón, D. A. Increasing topological diversity during computational "synthesis" of porous crystals: How and why. *CrystEngComm* **21**, 1653-1665 (2019).
42  Thompson, A. P. *et al.* Lammps - a flexible simulation tool for particle-based materials modeling at the atomic, meso, and continuum scales. *Comput. Phys. Commun.* **271**, 108171 (2022).
43  Coupry, D. E., Addicoat, M. A. & Heine, T. Extension of the universal force field for metal–organic frameworks. *J. Chem. Theory Comput.* **12**, 5215-5225 (2016).
44  Dubbeldam, D., Calero, S., Ellis, D. E. & Snurr, R. Q. Raspa: Molecular simulation software for adsorption and diffusion in flexible nanoporous materials. *Mol. Simul.* **42**, 81-101 (2016).
45  Wang, Y., Iyer, A., Chen, W. & Rondinelli, J. M. Featureless adaptive optimization accelerates functional electronic materials design. *Appl. Phys. Rev.* **7**, 041403 (2020).
46  Zhang, H., Chen, W., Iyer, A., Apley, D. W. & Chen, W. Uncertainty-aware mixed-variable machine learning for materials design. *Sci. Rep.* **12**, 19760 (2022).
47  Yerramilli, S., Iyer, A., Chen, W. & Apley, D. W. Fully bayesian inference for latent variable gaussian process models. *arXiv preprint arXiv:2211.02218* (2022).
48  Iyer, A., Yerramilli, S., Rondinelli, J. M., Apley, D. W. & Chen, W. Descriptor aided bayesian optimization for many-level qualitative variables with materials design applications. *J. Mech. Des.* **145** (2022).
49  Potoff, J. J. & Siepmann, J. I. Vapor–liquid equilibria of mixtures containing alkanes, carbon dioxide, and nitrogen. *AIChE J.* **47**, 1676-1682 (2001).
50  Rappe, A. K., Casewit, C. J., Colwell, K. S., Goddard, W. A., III & Skiff, W. M. Uff, a full periodic table force field for molecular mechanics and molecular dynamics simulations. *J. Am. Chem. Soc.* **114**, 10024-10035 (1992).
51  Kancharlapalli, S., Gopalan, A., Haranczyk, M. & Snurr, R. Q. Fast and accurate machine learning strategy for calculating partial atomic charges in metal–organic frameworks. *J. Chem. Theory Comput.* **17**, 3052-3064 (2021).
52  Zheng, J., Li, Z., Gao, L. & Jiang, G. A parameterized lower confidence bounding scheme for adaptive metamodel-based design optimization. *Eng. Comput.* **33**, 2165-2184 (2016).
53  Couckuyt, I., Deschrijver, D. & Dhaene, T. Fast calculation of multiobjective probability of improvement and expected improvement criteria for pareto optimization. *J. Global Optim.* **60**, 575-594 (2014).
54  Jones, D. R. A taxonomy of global optimization methods based on response surfaces. *J. Global Optim.* **21**, 345-383 (2001).

# Acknowledgements

W.C. acknowledges support from the Advanced Research Projects Agency-Energy (ARPA-E), U.S. Department of Energy, under Award Number DE-AR0001209, and National Science Foundation (NSF) Grant 1729743. R.Q.S. acknowledges support from the U.S. Department of Energy, Office of Basic Energy

Sciences, Division of Chemical Sciences, Geosciences and Biosciences under Award DE-FG02-17ER16362. T.D.P acknowledges computing support from Quest high performance computing facility at Northwestern University. T.D.P acknowledges computing support from the National Energy Research Scientific Computing Center (NERSC), a Department of Energy Office of Science User Facility located at Lawrence Berkeley National Laboratory, operated under Contract No. DE-AC02-05CH11231 using NERSC award BES-ERCAP0020094.

# Author Information

### Authors and Affiliations

**Department of Mechanical Engineering, Northwestern University, Evanston, IL 60208, USA**

Yigitcan Comlek, Wei Chen

**Department of Chemical and Biological Engineering, Northwestern University, Evanston, IL 60208, USA**

Thang Duc Pham, Randall Q. Snurr

### Author Contributions

Y.C. and W.C. conceived and designed the project. Y.C. led the collaboration, created the LVGP-MOBBO framework, performed the optimization, and analyzed the results. T.D.P. carried out the GCMC simulations and analyzed the results. Y.C. and T.D.P. wrote the manuscript. R.Q.S. and W.C. supervised the work. All authors (Y.C., T.D.P., R.Q.S., and W.C.) reviewed and edited the manuscript.

### Corresponding Author

Correspondence to Dr. Wei Chen (weichen@northwestern.edu)

## Ethics Declarations

**Competing interests**

R.Q.S. has a financial interest in the start-up company NuMat Technologies, which is seeking to commercialize metal−organic frameworks. The remaining authors declare no competing financial interests, and all authors declare no competing non-financial interests.

# Supplementary Information:

# Rapid Design of Top-Performing Metal-Organic Frameworks with Qualitative Representations of Building Blocks

Yigitcan Comlek[1], Thang Duc Pham[2], Randall Q. Snurr[2], Wei Chen[1]

[1]Department of Mechanical Engineering, Northwestern University, Evanston, IL, USA,

[2]Department of Chemical and Biological Engineering, Northwestern University, Evanston, IL, USA

# Contents

# Supplementary Figures

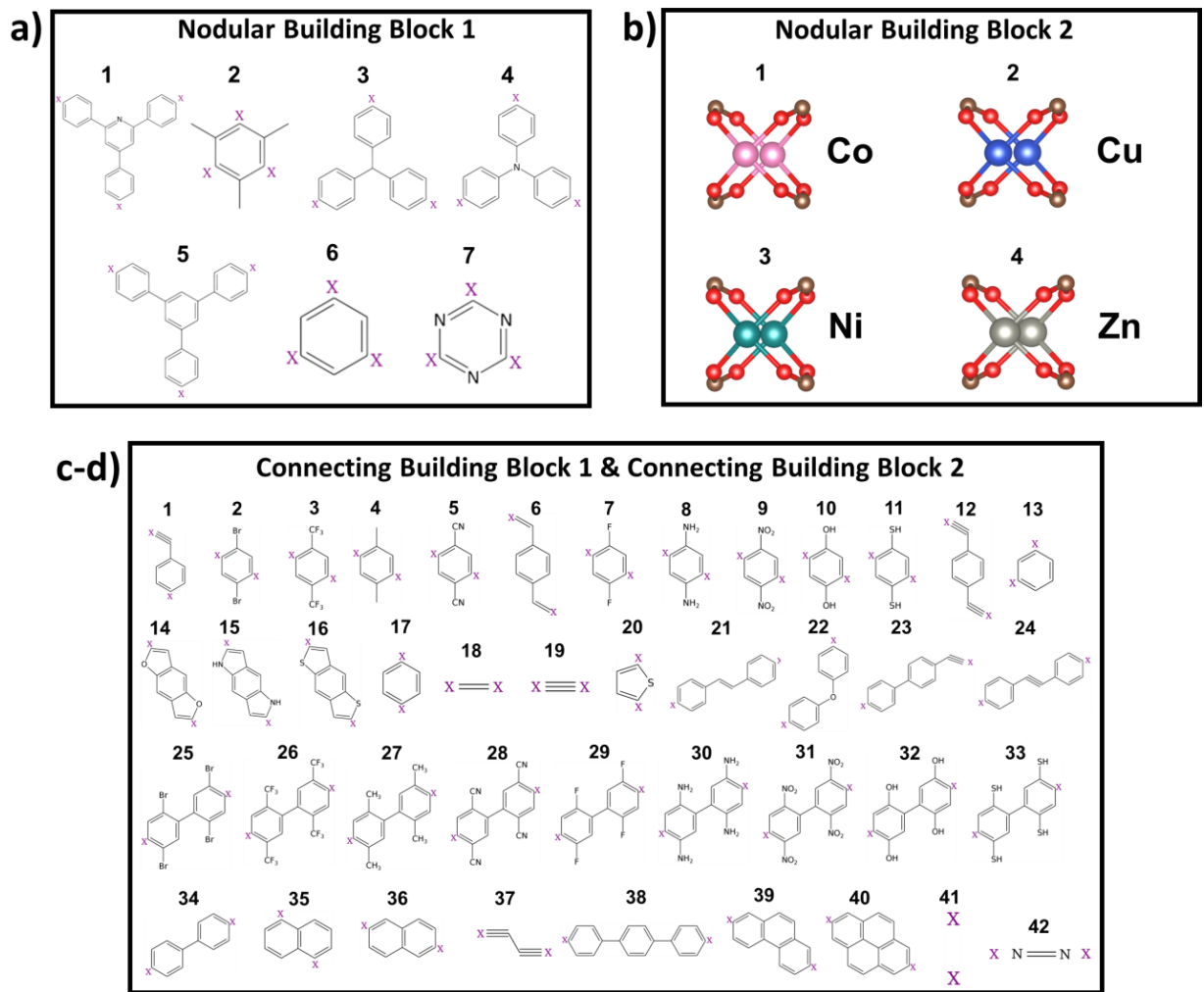

**Supplementary Figure 1.** The building blocks that make up the design spaces. There are (a) 7 Nodular Building Block 1 (Inorganic Node) and (b) 4 Nodular Building Block 2 (Organic Node) design options. (c-d) The Connecting Building Block 1 (c) and the Connecting Building Block 2 (d) have 41 and 42 different design options, respectively.

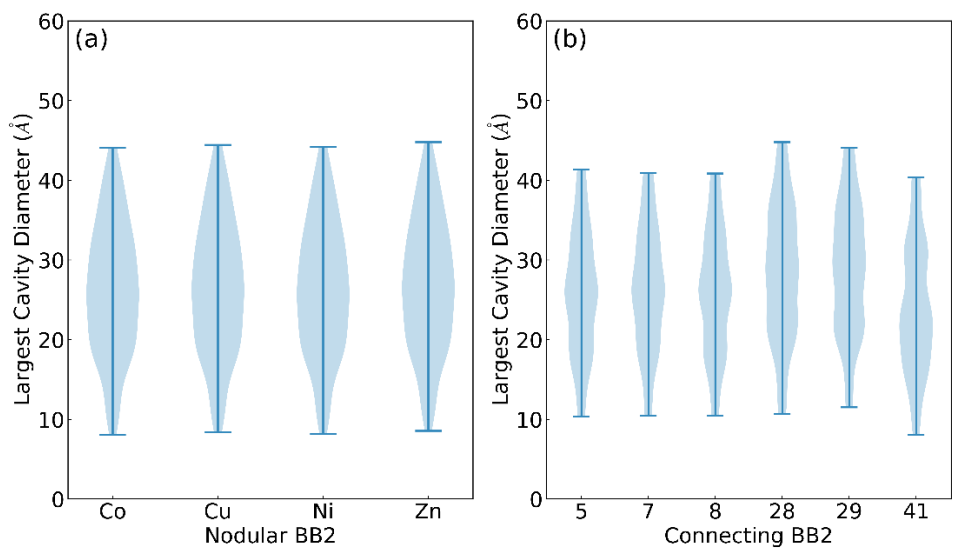

**Supplementary Figure 2**. The distribution of the largest cavity diameter (LCD) of 1001 MOFs in the Reduced Design Space (RDS) for different building blocks. LCD distribution for different (a) Nodular Building Block 2 and (b) Connecting Building Block 2 on the RDS.

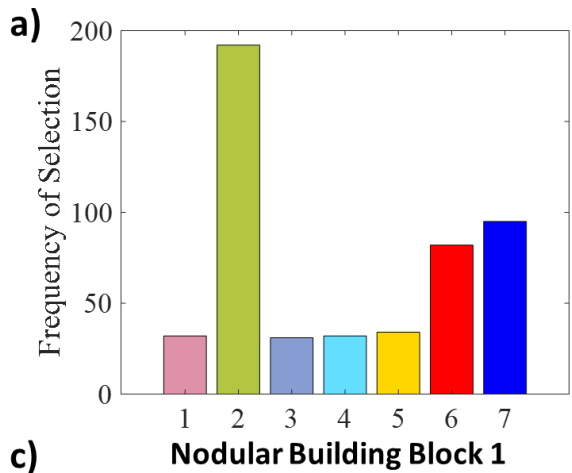
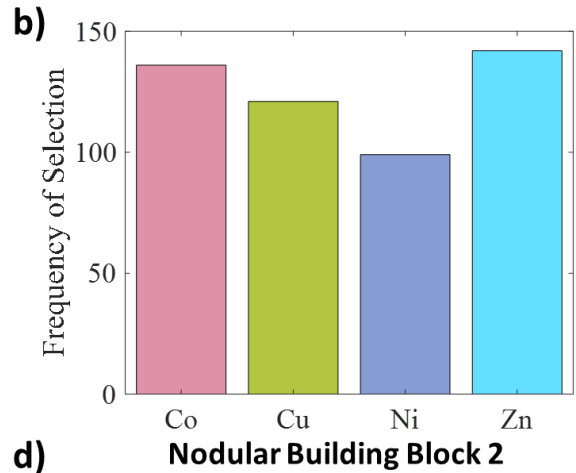
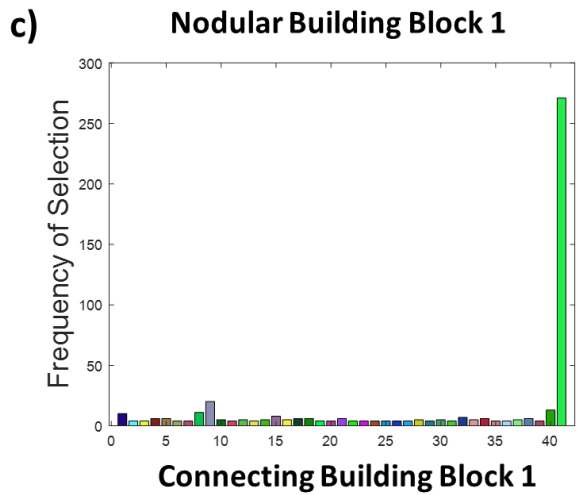
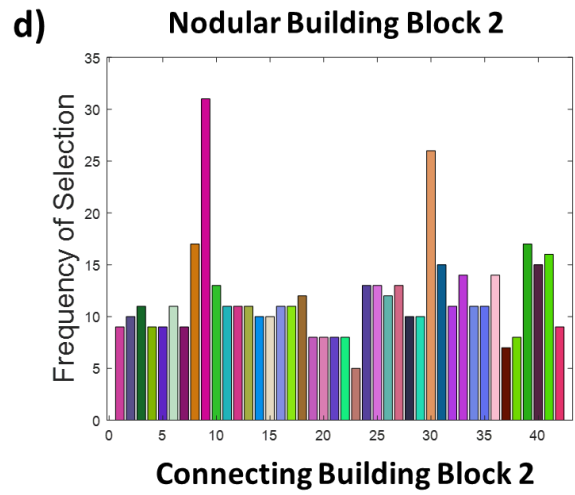

**Supplementary Figure 3.** Histogram of the building blocks selected after the 66 iterations of LVGP-MOBBO on the Entire Design Space (EDS). (a) Design decisions for the Nodular Building Block 1, (b) Nodular Building Block 2, (c) Connecting Building Block 1, and (d) Connecting Building Block 2.

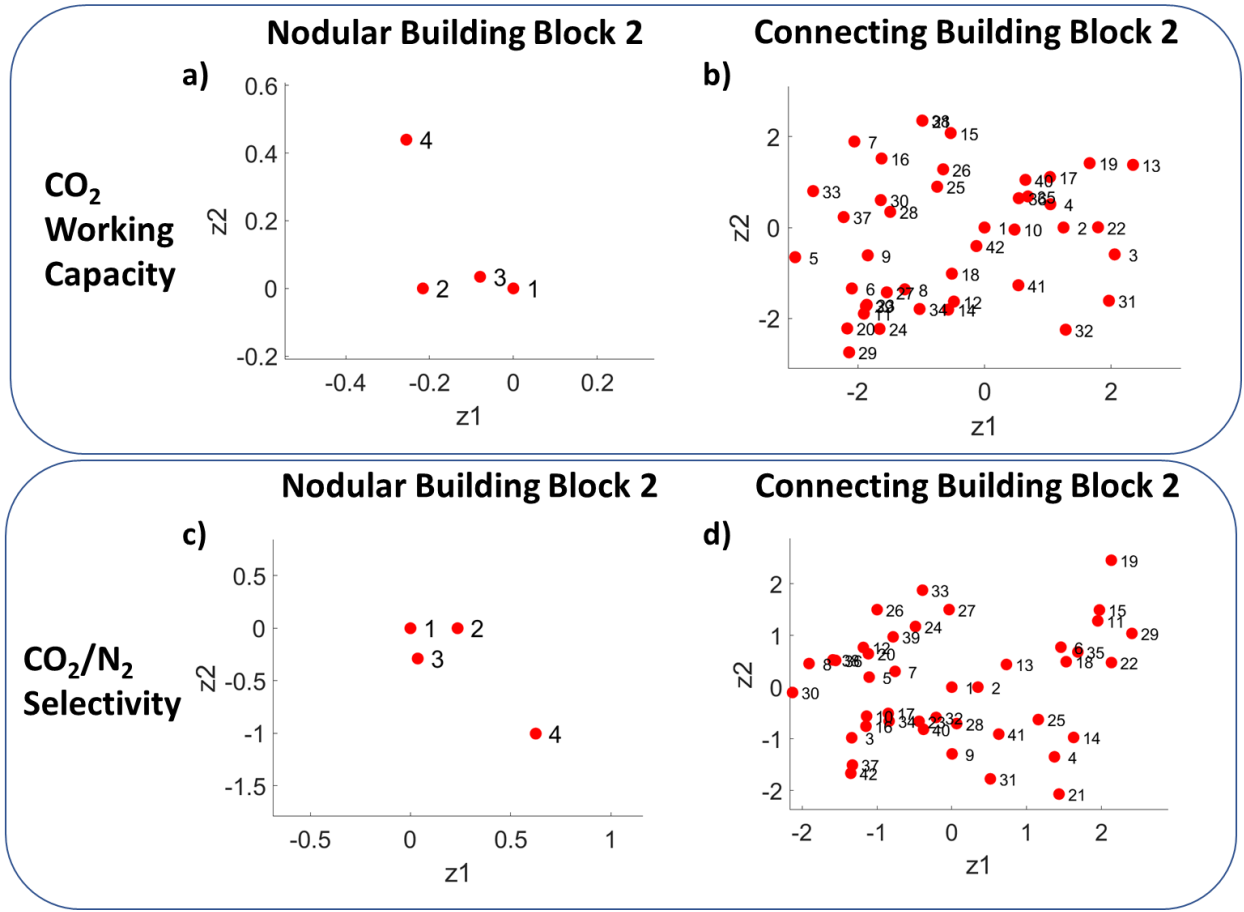

**Supplementary Figure 4.** Latent Variable Plots after the LVGP-MOBBO campaign on Large Building Block Design Space. Latent spaces of the (a) Nodular Building Block 2 and (b) Connecting Building Block 2 for the $CO_2$ working capacity. (c) Latent spaces for the Nodular Building Block 2 and (d) Connecting Building Block 2 for the $CO_2/N_2$ selectivity.

# Supplementary Table

**Supplementary Table 1.** Force field parameters for the metal in the metal nodes. Lennard Jones parameters and the range of partial charges of the metal in the inorganic building blocks.

| Atoms | $\sigma$ (Å) | $\varepsilon/k_B$ (K) | Range of Partial Charges |
|---|---|---|---|
| Co | 2.56 | 7.040 | 0.73 - 1.08 |
| Cu | 2.52 | 3.114 | 0.72 - 1.04 |
| Ni | 2.52 | 7.550 | 0.71 – 1.05 |
| Zn | 2.46 | 62.40 | 0.68 – 1.00 |